\documentclass[12pt]{iopart}

\begin{document}
\newcommand{\be}{\begin{equation}}
\newcommand{\ee}{\end{equation}}
\newcommand{\ben}{\begin{eqnarray}}
\newcommand{\een}{\end{eqnarray}}
\newcommand{\n}{\nonumber  }
\newcommand{\nn}{\nonumber \\ }
\newcommand{\nd}{\noindent}
\newcommand{\p}{\partial}
\title{Parameter-free ansatz for inferring ground state wave functions  of even potentials}

\author{S.P. Flego$^1$, 
 A. Plastino$^{2,\,4, \,5}$, 
   A.R. Plastino$^{3,\,4}$}

\address{$^{1}$Universidad Nacional de La Plata, Facultad de Ingenier\'{\i}a, \'Area Departamental de Ciencias B\'asicas,
 1900 La Plata, Argentina \\  $^{2}$Universidad Nacional de La Plata, Instituto
de F\'{\i}sica (IFLP-CCT-CONICET), C.C. 727, 1900 La Plata, Argentina \\
$^{3}$CREG-Universidad Nacional de La Plata-CONICET, C.C. 727, 1900 La Plata, Argentina\\
$^{4}$Instituto Carlos I de Fisica Teorica y Computacional and
Departamento de Fisica Atomica, Molecular y Nuclear, Universidad
de Granada, Granada, Spain \\$^{5}$  Universitat de les Illes
Balears and IFISC-CSIC, 07122 Palma de Mallorca, Spain}
\ead{angeloplastino@gmail.com\hskip 3mm (corresponding author)}
\begin{abstract}
\nd  
Schr\"odinger's equation (SE) and the information-optimizing
principle based on Fisher's information measure (FIM) are intimately linked, which entails the existence of a Legendre transform
structure underlying the SE.  
In this comunication we show that the existence of such an structure allows, via the virial theorem,  for the formulation 
of a parameter-free ground state's  SE-ansatz for a rather large family of potentials. The parameter-free nature of the ansatz derives from 
the structural information it incorporates through its Legendre properties.



\end{abstract}
\nd 
\maketitle

\section{ Introduction}

\nd Few quantum-mechanical models admit of exact solutions. 
Approximations of diverse type constitute the hard-core of the armory at the disposal of the quantum-practitioner. 
Since the 60's, hypervirial theorems have been gainfully incorporated to the pertinent arsenal \cite{castro,wewe}. We revisit here the subject
 in an information-theory context, via Fisher's information measure (FIM) with emphasis on  i) its Legendre properties and ii) its relation with the virial theorem.

\nd Remark that the notion of using a small set of relevant expectation values so as to describe the main 
properties of physical systems may be considered the leit-motiv of statistical mechanics \cite{brillu}. 
 Developments based upon Jaynes' maximum entropy principle constitute a pillar of our present understanding of the discipline \cite{katz}. 
  This type of ideas has also been fruitfuly invoked for obtaining the probability distribution associated to pure 
quantum states via Shannon's entropy (see for instance, \cite{pp} and references therein).
  In such a spirit,    Fisher information, the local counterpart of the global Shannon quantifier \cite{frieden2}, first introduced for statistical estimation purposes \cite{frieden2}.
 has been shown to be quite useful for the variational characterization of quantal equations of motion \cite{luo}.
 In particular, it is well-known that a strong link exists between  Fisher's
information measure  (FIM) $I$  and Schr\"odinger's
wave equation (SE)
\cite{pla7,flego,reginatto,HF-TV-RR,ArX1Univ,ArX2Alpha,ArX3}. Such connection is based upon the fact that a
constrained Fisher-minimization leads to a SE-like equation
\cite{frieden2,pla7,flego,reginatto,HF-TV-RR,ArX1Univ,ArX2Alpha,ArX3}.
In turn, this guarantees the existence of intriguing relationships
between various  quantum quantities reminiscent of the ones that characterize thermodynamics due to its Legendre-invariance
structure \cite{pla7,flego}.
Interestingly enough,  SE-consequences such as the
Hellmann-Feynman and the Virial theorems can be re-interpreted in
terms of thermodynamics' Legendre reciprocity relations
\cite{ArX1Univ,HF-TV-RR}, a fact suggesting that a
Legendre-transform structure underlies the non-relativistic
Schr\"odinger equation. As a consequence, the possible
energy-eigenvalues become constrained by such
structure in a rather unsuspected way
\cite{HF-TV-RR,ArX1Univ,ArX2Alpha,ArX3}, which allows one to
obtain a first-order differential equation, unrelated to
Schr\"oedinger's equation \cite{ArX2Alpha,ArX3},  that energy
eigenvalues must necessarily satisfy. The predictive power of that
 equation was explored in \cite{AHO4},
where the formalism was applied to the quantum anharmonic oscillator.
 Exploring further interesting properties
 of this ``quantal-Legendre'' structure will occupy us below. As a result, it will be seen that, as a 
direct consequence of the Legendre-symmetry that underlies the 
connection between Fisher's measure and Schr\"oedinger's equation one immediately encounters an elegant expression 
for an ansatz, in terms of quadratures, of the ground state (gs) wave function of a rather wide category of potential functions. 

\section{ Basic ideas} 

\nd A special, and particularly useful FIM-expression (not the most general one)  is to be quoted. Let $x$ be a stochastic
variable  and $f(x)= \psi(x)^2$ the
probability density function (PDF) for this variable. Then $I$ reads \cite{frieden2}

 \ben \label{eq.1-2}  I = \, \int f(x)
\left(\frac{\partial \ln{f(x)} }{\partial x} \right)^2 dx =4 \int\,dx\, [\nabla \psi(x)]^2 ~;\hspace{0.5cm}\,\,\,f=\psi^2. \een
Focus attention now a system that is specified by a set of $M$
physical parameters $\mu_k$. We can write $\mu_k = \langle
A_{k}\rangle,$ with  $A_{k}= A_{k}(x).$ The set of
$\mu_{k}$-values is to be regarded as our prior knowledge (available empirical information). Again,
the probability distribution function (PDF) is called $f(x)$. Then, \be
\label{eq.1-3} \langle A_{k}\rangle\,=\,\int ~dx ~A_{k}(x) ~f(x),
\hspace{0.5cm} k=1,\dots ,M. \ee It can be shown
(see  \cite{pla7,reginatto}) that the {\it physically
relevant} PDF $f(x)$  minimizes  FIM subject to
the prior conditions and the normalization condition.
Normalization entails $\int dx  f(x) = 1,$  and, consequently, our
Fisher-based extremization problem becomes \be
\label{eq.1-4}\delta \left( I - \alpha \int ~dx ~f(x) -
\sum_{k=1}^M~\lambda_k\int ~dx ~A_{k}(x)~f(x)\right) = ~0, \ee
with $(M+1)$ Lagrange multipliers
$\lambda_k$ ($\lambda_0=\alpha$). The reader is referred to Ref. \cite{pla7} for
the details of how to go  from (\ref{eq.1-4}) to a Schr\"odinger's
equation (SE) that yields the desired PDF in terms of the
amplitude $\psi(x)$. This SE is of the
form
 \be \label{eq.1-5} \left[-~\frac{1}{2}~\frac{\partial^2 ~}{\partial x^2} +U(x)\right]\psi ~= ~ \frac{\alpha}{8}~ \psi,\hspace{1.3cm} U(x)
=~-\frac{1}{8}~\sum_{k=1}^{M}\,\lambda_{k}~ A_{k}(x),\ee
and is to be interpreted as the (real) Schr\"odinger
equation (SE) for a particle of unit mass ($\hbar=1$) moving in the
effective, ``information-related pseudo-potential" $U(x)$ \cite{pla7}
 in which the normalization-Lagrange multiplier ($\alpha /8$) plays the role
of an energy eigenvalue. The  $\lambda_k$ are fixed by
recourse to the available prior information. For
one-dimensional scenarios,   $\psi(x)$ is real \cite{richard} and
\ben \label{eq.1-7}  I =   \, \int  \psi^2
\left(\frac{\partial \ln{\psi^2} }{\partial x} \right)^2 dx\,=\, 4
\int  \left(\frac{\partial \psi }{\partial x} \right)^2 dx
 =\, -4 \int  \psi \frac{\partial^2 ~}{\partial x^2} \psi~dx
 \een
\nd so from (\ref{eq.1-5}) one finds a simple and convenient $I-$expression
\ben \label{eq.1-12} I=\,\alpha
 + \sum_{k=1}^M~\lambda_k\left\langle A_k\right\rangle. \een

\vskip 2mm 

 \nd   {\bf Legendre structure}

\vskip 2mm

\nd The connection between the variational solutions $f$  and
thermodynamics was established in Refs. \cite{pla7} and \cite{flego} in the guise of
typical Legendre reciprocity relations.
These constitute thermodynamics' essential formal ingredient \cite{deslog}
and were re-derived \`a la Fisher in \cite{pla7} by recasting (\ref{eq.1-12})
in a fashion that emphasizes the role of the relevant independent
variables,
\ben \label{eq.1-13a} I(\langle
A_1\rangle,\ldots,\langle A_M\rangle) \,=\,\alpha
  + \sum_{k=1}^M~\lambda_k \langle A_k\rangle. \een
 Obviously, the Legendre transform main goal is that of  changing the identity
of our relevant independent  variables. As for the normalization multiplier $\alpha$, that plays the role of an energy-eigenvalue in Eq. (\ref{eq.1-5} ), we have
\be \label{eq.1-13b}\alpha(\lambda_1,\ldots,\lambda_M)= I-
\sum_{k=1}^M~\lambda_k\left\langle A_k\right\rangle. \ee
After these preliminaries we straightforwardly  encounter the three reciprocity relations
 \cite{pla7}
\be \label{RR-1} \frac{\partial \alpha}{\partial \lambda_{k}}= -
\langle A_k\rangle ~; \hspace{1.cm}  \frac{\partial I }{\partial
\langle A_k \rangle}\,=\,\lambda_k  ~ ;\hspace{1.cm}
\frac{\partial I}{\partial \lambda_{i}}=\sum_{k}^{M} \lambda_{k}
 \frac{\partial \langle A_k \rangle}{\partial \lambda_{i}},\ee
the last one being a generalized Fisher-Euler theorem.

\section{ Fisher measure and quantum mechanical connection}

\nd Since the potential function $U(x)$ belongs to $\mathcal{L}_2$,
 it admits of a series expansion in the basis
$x,\,x^2\,x^3,\,etc.$ \cite{greiner}. The $A_k(x)$ themselves
belong to $\mathcal{L}_2$ as well, and can also  be series-expanded
in similar fashion. This enables us to base our future
considerations on the assumption that the a priori knowledge
refers to moments $x^k$ of the independent variable, i.e., $
\langle A_k \rangle~=~ \langle x^k \rangle $, and that one
possesses information about  $M$ of these moments
 $\langle x^k \rangle$. Our ``information" potential $U$
  thus reads
\be \label{virial-5} U(x)=  -~ \frac{1}{8} \sum_k \,\lambda_k\,
x^k. \ee  {\it We will assume that the first $M$ terms of the
above series yield a satisfactory
 representation of} $~U(x)$. Consequently, the Lagrange multipliers are
 identified with  U(x)'s series-expansion's coefficients.

\nd In a Schr\"odinger-scenario the {\it virial theorem} states that \cite{HF-TV-RR}
 \ben \label{virial-6}
\left\langle \frac{\partial^2~}{\partial x^2}\right\rangle = -~
\left\langle x ~ \frac{\partial ~}{\partial x} U(x)\right\rangle ~
= ~ \frac{1}{8}~\sum_{k=1}^{M}\, k \,\lambda_{k}~\langle x^k
\rangle ~, \een and thus, from (\ref{eq.1-7}) and (\ref{virial-6})
a useful, virial-related expression for Fisher's information
measure can be arrived at \cite{HF-TV-RR}
\ben \label{virial-7} I\,
=\, -~ ~\sum_{k=1}^{M}\, \frac{k}{2} \,\lambda_{k}~\langle
x^{k}\rangle, \een
 $I$ is explicit function of the $M$ physical parameters $\langle
x^{k}\rangle$. Eq. (\ref{virial-7}) encodes the information provided by the virial theorem \cite{ArX1Univ,HF-TV-RR}.

\vspace{0.5cm}

\nd  Interestingly enough, the reciprocity relations (RR)
(\ref{RR-1}) can be re-derived on a strictly pure quantum
mechanical basis \cite{HF-TV-RR}, starting from
the quantum Virial theorem [which leads to Eq.
(\ref{virial-7}) ] plus information
provided by the quantum Hellmann-Feynman theorem. This fact strongly suggests that a Legendre structure underlays  the one-dimensional
Schr\"oedinger equation \cite{HF-TV-RR}. Thus, with $\langle
A_k\rangle=\langle x^k\rangle$, our ``new"  reciprocity relations
are given by \be \label{RR-q} \frac{\partial \alpha}{\partial
\lambda_{k}}= - \langle x^k\rangle ~; \hspace{1.cm} \frac{\partial
I }{\partial \langle x^k \rangle}\,=\,\lambda_k  ~ ;\hspace{1.cm}
\frac{\partial I}{\partial \lambda_{i}}=\sum_{k}^{M} \lambda_{k}
 \frac{\partial \langle x^k \rangle}{\partial \lambda_{i}},\ee
 FIM expresses a relation between the independent variables or
control variables (the prior information) and $I$. Such
information is encoded into the functional form  $I=I(\langle x^1
\rangle, ... , \langle x^M \rangle  )$. For later convenience, we
will also denote such a relation or encoding-process as $\{I,\langle x^k
\rangle \}$. We see that the Legendre transform FIM-structure
involves both eigenvalues of the ``information-Hamiltonian" and
Lagrange multipliers. Information is encoded in
 $I$ via these Lagrange multipliers, i.e., ${\alpha}={\alpha}(\lambda_1, ... \lambda_M),$ $
{\rm together \,\, with \,\, a  \,\, bijection} \,\,\,\{I,\langle
x^k \rangle \}  \hspace{0.6cm} \longleftrightarrow \hspace{0.6cm}
\{{\alpha}, \lambda_k \}. \label{RR-3}$

\nd {\bf In a  $\left\{I, \langle {x}^{k}\rangle \right\}$ -
scenario}, the $\lambda_k$ are  functions dependent on the $\langle
 {x}^{k}\rangle$-values.
As  shown in \cite{ArX1Univ}, substituting the RR given by
(\ref{RR-q}) in (\ref{virial-7}) one is led to a {\it linear,
partial differential equations (PDE)} for $I$, \ben \label{gov-1}
\lambda_k \,=\, \frac{\partial I }{\partial \langle x^k \rangle}
\hspace{1.cm}\longrightarrow \hspace{1.cm}  I \,  =\,
-~\sum_{k=1}^{M}\,\frac{k}{2} ~\langle x^{k}\rangle ~
\frac{\partial I }{\partial \langle x^k \rangle}\,. \een and a
complete solution is given by
\ben \label{gov-9} {I}(\langle {x}^1
\rangle, ... , \langle {x}^M \rangle ) =~
 \sum_{k=1}^{M}~C_k~ ~{ \left| \langle
{x}^{k}\rangle \right|^{- {2}/{k}}}~,\een
 where $C_k$ are positive real numbers (integration constants).
The $I$ - domain is ${\it D}_I=\left\{(\langle {x}^1 \rangle, ...
, \langle {x}^M \rangle) / \langle {x}^k \rangle~\in ~\Re_o
\right\}$.  Eq. (\ref{gov-9}) states that for $\langle {x}^k
\rangle >0$, $I$ is a monotonically decreasing function of $ \langle
x^k \rangle$, and as one expects from a ``good'' information
measure \cite{frieden2}, $I$ is a convex function. We may obtain
$\lambda_k$ from the reciprocity relations (\ref{RR-q}). For
$\langle x^k \rangle ~ > ~ 0 ~$ one gets,
\ben \label{prop-2-RR} \lambda_k ~=~ \frac{\partial I}{\partial \langle
x^k \rangle} ~=~-~ \frac{2}{k}~C_k~ ~{ \langle {x}^{k}\rangle^{-~
(2+k)/k}}~<~0~. \een and then, using (\ref{eq.1-12}), we obtain
the $\alpha$ - normalization Lagrange multiplier.
 For a discussion on how to obtain the reference quantities $C_k$ see
 \cite{AHO4}.

\nd The general solution for the $I$ - PDE does exist and its
uniqueness has been demonstrated via an analysis of the associated
Cauchy problem \cite{ArX1Univ}. Thus, Eq. (\ref{gov-9}) implies
what seems to be a kind of ``universal'' prescription, a  linear
PDE that any variationally (with constraints) obtained FIM must
necessarily comply with.

\section{ Present results}

\subsection{Inferring the PDF for even potentials}
\nd  For even informational potentials good SWE-ansatz can be formulated via probability distribution functions (PDF) that satisfy the virial theorem.
The potentials are of the form 
\ben \label{pdf-1} U(x)=-\frac{1}{8}\sum_{k=1}^{M} \lambda_{2k} ~ x^{2k},\een
 and the ansatz can be straightforwardly derived from  (\ref{eq.1-2}) and (\ref{virial-6}).
This constitutes our main present result. The procedure is as follows. Begin with
the Fisher measure $I$, ``virially'' expressed as
\ben \label{pdf-2}
I~=~- 4 \left\langle \frac{\partial^2~}{\partial x^2}\right\rangle = 4
\left\langle x ~ \frac{\partial ~}{\partial x} U(x)\right\rangle ~
\hspace{0.5cm}\longrightarrow \hspace{0.5cm}I~= ~- \left\langle ~\sum_{k=1}^{M}\, k \,\lambda_{2k}~x^{2k}
\right\rangle ~, \een
\nd which, in the Fisher-scenario, can obviouly be written as

\ben \label{pdf-3}
\int~ dx~ f(x) \left(\frac{\partial \ln{f(x)} }{\partial x} \right)^2 ~= ~
\, -~ \int~dx~f(x)~\sum_{k=1}^{M}\, k \,\lambda_{2k}~ x^{2k}~,
\een
or
\ben \label{pdf-4}
\int~ dx~ f(x) \left[\left(\frac{\partial \ln{f(x)} }{\partial x} \right)^2
\, +~ ~\sum_{k=1}^{M}\, k \,\lambda_{2k}~ x^{2k}\right]~= ~0.
\een

\nd We devise an ansatz $f_A$ that by construction verifies (\ref{pdf-4}). We merely require fulfillment of 
\ben \label{pdf-5}
\left(\frac{\partial \ln{f_A(x)} }{\partial x} \right)^2
\, +~ ~\sum_{k=1}^{M}\, k \,\lambda_{2k}~ x^{2k}~= ~0.
\een

\nd Clearly, we inmediatly  obtain,
\ben \label{pdf-6}
\left(\frac{\partial \ln{f_A(x)} }{\partial x} \right)^2  &=&
\,- \sum_{k=1}^{M}\, k \, \lambda_{2k}~ x^{2k}~,\een
that leads to
\ben \label{pdf-7}
f_A(x)~ =~exp \left\{-\int{ dx~\sqrt{
\, - \sum_{k=1}^{M}\, k \,\lambda_{2k}~ x^{2k}}}\right\}~,
\een
where the minus sign in  the exponential argument was chosen
 so as to enforce  the condition that $f(x)~ \stackrel{x\rightarrow\pm\infty}{\longrightarrow}~0$.
 Eq. (\ref{pdf-7}) provides us with a nice, rather general and virially motivated ansatz. Is it good enough  for dealing with the SWE?.
We look for an answer below.

\subsection{Harmonic oscillator (HO)}

\nd It is obligatory to start our investigation  with reference  to
 the harmonic oscillator.
 \nd One assumes that the  prior Fisher-information is given by
 \ben \label{ex-HO-1}  \langle x^2\rangle ~ = ~\frac{1}{2\omega}~.\een
\nd The pertinent FIM  can now be obtained by using (\ref{gov-9}),
\ben  I ~ = ~I({\langle x^2\rangle})~= ~C_2~
\langle x^2 \rangle ^{- 1} ~,\n \een
 which saturates  the Cramer-Rao bound \cite{frieden2} when $C_2=1$,
\ben \label{ex-HO-2} I~\langle x^2 \rangle ~=~C_2~=~1 \hspace{1.cm}\Longrightarrow \hspace{1.cm}I ~ =  ~\langle x^2 \rangle ^{- 1} ~.  \een

\nd The pertinent Lagrange multiplier can be obtained by recourse to the reciprocity relations (\ref{RR-1}) and (\ref{ex-HO-2}),
\ben \label{ex-HO-3} \lambda_2 \,=~ \frac{\partial I }{\partial \langle x^2 \rangle}
\,=\, -~ \langle x^2 \rangle ^{- 2}~. \een
\nd The prior-knowledge (\ref{ex-HO-1}) is encoded into the FIM (\ref{ex-HO-2}),
and the Lagrange multiplier $\lambda_2$ (\ref{ex-HO-3}),
\ben \label{ex-HO-4} I ~ = ~\langle x^2\rangle~^{-1}= ~ 2 \omega \,;
 \hspace{1.5cm}  \lambda_2 ~=~-~ \langle x^2 \rangle
~^{- 2}~= ~ - ~ 4\omega^2~. \een
and the $\alpha -$value is gotten from (\ref{eq.1-13b}),
\ben \label{ex-HO-5} \alpha  ~=~ I-~\lambda_2 ~ \langle x^2\rangle
=~4~\omega. \een

\nd Our ansatz-PDF can be extracted from (\ref{pdf-7}) as follows
\ben \label{ex-HO-6}
f(x)~ =~exp \left\{-\int{ dx~\sqrt{
\,  \,4\omega^2~ x^{2}}}\right\}~=~N~exp \left\{-\,\omega x^{2}\right\}~,
\een
with,
\ben \label{ex-HO-7}
\int ~f(x)~dx =~1 \hspace{1.cm}\longrightarrow\hspace{1.cm}N~=
 \left(\frac{\omega}{\pi}\right)^{1/2}~,
\een
the exact result.

\section{ Ground state eigenfunction of the general,\\
\hspace*{0.5cm} even-anharmonic oscillator}

\nd We  outline here  the  methodology  for constructing the ground state ansatz for an anharmonic oscillator  of the form 
(we shall herefrom omit the subscript A)

\ben \label{AHO-1}
\left[-\frac{1}{2}\frac{d^2}{dx^2}+ \sum a_{2k} x^{2k} \right]\psi(x)~=~ E\psi(x)\een
\nd According to \cite{ArX2Alpha,ArX3}, we can ascribe to (\ref{AHO4-1}) a Fisher measure and effect then the following identifications:
\ben \label{AHO-2}~ \alpha = 8 E \,,~\hspace{1.cm}
  \lambda_{2k}= -8~a_{2k} \,,\hspace{1.cm}~ f(x)=\psi^2(x).\een

\nd Accordingly, we get our ansatz by substituting into (\ref{pdf-7}) the quantities given by (\ref{AHO-2}).
\ben \label{AHO-3}
\psi(x)~ =~exp \left\{-\frac{1}{2}\int{ dx~\sqrt{
\,  \sum_{k=1}^{M}\,8~ k \,a_{2k}~ x^{2k}}}\right\}~,
\een
\nd As an illustration of  the procedure, we deal below  with the quartic anharmonic oscillator.

\subsection{Quartic anharmonic oscillator}

\nd The Schr\"odinger equation for a particle of unit mass in a
quartic anharmonic potential reads,
\be \label{AHO4-1}
\left[-~\frac{1}{2}~\frac{\partial^2~}{\partial x^2}
+~\frac{1}{2}~\omega^2~ x^2~+ \frac{1}{2}~\lambda ~x^4 \right]\psi~= ~ E~
\psi, \ee
\nd where $\lambda$ is the anharmonicity constant.
Expression (\ref{AHO-3}) takes the form
\ben \label{AHO4-2}
\psi(x)~=~exp \left\{-\frac{1}{2}\int{ ~\sqrt{
\, 4 \,\omega^2~ x^{2}+8 \,\lambda~ x^{4}}}~dx\right\}.\n\een
\nd Now,  from an elemental integration, we obtain the desired eigenfunction
\ben \label{AHO4-3}
\psi(x)& =&~N~exp \left\{\frac{\omega^3}{6\lambda}\left[1-
~ \left(1+ \frac{2 \lambda}{\omega^2} ~ x^{2}\right)^{3/2}\right]~\right\}~,
\een
where $N$ is the normalization constant. 

\vspace{0.5cm}

\nd When $\lambda \rightarrow 0$ one re-obtains the Gaussian form,
\ben
\lim_{\lambda \rightarrow 0} \psi(x)~ = ~\psi_{HO}~=~N~exp \left(-\,\omega x^{2}\right)~,
\een

\nd and, when $\omega \rightarrow 0$ the  pure anharmonic oscillator eigenfunction is given by, 
\ben
\lim_{\omega \rightarrow 0} \psi(x) =~\psi_{PAO}~=~ N~exp \left(-\,\frac{\sqrt{2\lambda}}{3}~|x|^{3}\right)~,
\een

\nd Once we have at our dispossal the anzsatz gs-eigengenfunction, we  obtain the corresponding eigenvalues following one of the two procedures.

\hspace{0.6cm}

\nd {\it Schr\"oedinger procedure:}

\ben \label{proc-1}
E& \approx &\langle \psi | H | \psi \rangle =\int~dx~\psi(x)\left[-~\frac{1}{2}~\frac{\partial^2~}{\partial x^2}
+~\frac{1}{2}~\omega^2~ x^2~+ \frac{1}{2}~\lambda ~x^4 \right]\psi(x)~=\n \\
&=&\int~dx~\psi(x)\left[
\frac{\omega}{2}\left(1+\frac{2 \lambda}{\omega^2}~x^2\right)^{1/2}+
\frac{\lambda}{\omega}~x^2~\left(1+\frac{2 \lambda}{\omega^2}~x^2\right)^{-1/2}-\frac{\lambda}{2}~x^4\right]\psi(x)~.~\hspace{0.7cm}
\een

\vspace{0.6cm}

\nd{\it Fisher procedure:}

\vspace{0.3cm}

\nd From (\ref{eq.1-12}) and (\ref{virial-7}), with $\lambda_2=-4\omega^2~,~\lambda_4=-4\lambda$, we have
\ben \label{proc-2} \alpha~=\,I - \sum_{k=1}^M~\lambda_k\left\langle x^k\right\rangle ~=~-\sum_{k=1}^{M}\,\left( \frac{k}{2}+1\right) \,\lambda_{k}~\langle
x^{k}\rangle ~= ~8 ~\omega^2 \,\langle x^{2}\rangle + 12~ \lambda~\langle
x^{4}\rangle ~, \een
Evaluating the moments with the anzsatz function, we have
\ben
\langle x^{p}\rangle_{_A}~\approx~\int dx ~x^p~f(x)~=~\int dx ~x^p~\psi^2(x)
\een
and, accordingly,
\ben
E=\frac{\alpha}{8}~\approx~\omega^2 \,\langle x^{2}\rangle_{_A} + \frac{3}{2}~ \lambda~\langle
x^{4}\rangle_{_A} ~, \een
\nd We  determine
$E$ {\it without passing first through a Schr\"odinger equation}, which is a nice 
aspect of the present approach. The question for the suitability of our ansatz is answered by looking at the Table below.

\vspace{0.7cm}

\small
\nd  {\bf Table}: 

SE-ground-state eigenvalues 
(\ref{AHO4-1}) for $\omega=1$ and several
values of the anharmonicity-constant $\lambda$.

\vspace{0.3cm}

\nd \begin{minipage}{8.20cm}
 The values of the second column correspond to those one
finds in the literature,  obtained via a numerical approach to
the SE. {\it These} results, in turn,  are
nicely reproduced by some interesting theoretical approaches that,
however, need to introduce and adjust some empirical constants
\cite{Banerjee}. Our ansatz-values, in the third column, are obtained by
 a parameter-free procedure. The
fourth column displays  the associated Cramer-Rao bound, which is almost saturated in all instances.
\rm
\end{minipage}
\begin{minipage}[t]{8.0cm}
\footnotesize
\begin{flushright}
\begin{tabular}{||l||c||c|c||}
\hline
\hline
\hspace{0.4cm}$~\lambda~$ &~$~E_{num}~$~ & $E$ &
$I~\langle {x}^{2}\rangle$ \\
\hline
\hline
~0.0001&~0.50003749&~0.50003749&~1.000000015\\
~0.001&~ 0.50037435&~0.50037444&~1.000001477\\
~0.01&~  0.50368684&~0.50369509&~1.000129847\\
~0.1&~   0.53264275&~0.53305374&~1.000129847\\
~1&~     0.69617582&~0.70188134&~1.046344179\\
~10&~    1.22458704&~1.25080186&~1.099588057\\
~100&~   2.49970877&~2.57093830&~1.123451126\\
~1000&~  5.31989436&~5.48276171&~1.130099216\\
\hline
\hline
\end{tabular}
\end{flushright}
\rm

\end{minipage}

\vspace{0.5cm}

\section{ Conclusions}

\nd The link Schr\"oedinger equation - Fisher measure has been employed so as to infer, via the pertinent reciprocity relations, 
 a parameter-free  ground state ansatz wave function for a rather ample family of even potentials, of the form

  \be  U(x)=  \sum a_{2k} x^{2k},  \ee
in terms of the coefficients $a_{2k}$. Its parameter-free character notwithstanding, our ansatz provides good results, as evidenced by  the examples  here examined. 
It incorporates only the knowledge of the virial theorem, via the Legendre-symmetry that underlies the 
connection between Fisher's measure and Schr\"oedinger equation.  One may again speak here of the  power of symmetry considerations.
in devising physical treatments.

\vspace{0.5cm}


\nd {\bf Acknowledgments-} This work was partially supported by the
Projects FQM-2445 and FQM-207 of the Junta de Andalucia
(Spain, EU).

\vspace{0.5cm}


\end{document}